\documentclass[conference,a4paper]{IEEEtran}

\usepackage[utf8]{inputenc} 
\usepackage[T1]{fontenc}
\usepackage{url}              
\usepackage{cite}             

\usepackage[cmex10]{amsmath}  
\usepackage{mleftright}       
\mleftright                   

\usepackage{graphicx}         
\usepackage{booktabs}         

\usepackage{amsmath,amssymb,amsfonts}
\usepackage[colorlinks,citecolor=blue,urlcolor=blue,filecolor=blue]{hyperref}

\usepackage{wrapfig}
\usepackage{caption}
\usepackage{subcaption}
\usepackage{booktabs}   

\usepackage{amsthm}

\newtheorem{theorem}{Theorem}

\begin{document}

\title{Context-tree weighting for real-valued time series:\\
Bayesian inference with hierarchical mixture models}

\author{\IEEEauthorblockN{Ioannis Papageorgiou}
\IEEEauthorblockA{\textit{University of Cambridge} \\
\texttt{ip307@cam.ac.uk}}
\and
\IEEEauthorblockN{Ioannis Kontoyiannis}
\IEEEauthorblockA{\textit{University of Cambridge} \\
\texttt{yiannis@maths.cam.ac.uk}}
}

\maketitle

\begin{abstract}
Real-valued time series are ubiquitous in the sciences and engineering. 
In this work, a general, hierarchical Bayesian modelling 
framework is developed for building mixture models for times series.
This development is based, in part, on the use of 
{\em context trees}, and it includes 
a collection of effective algorithmic 
tools for learning and inference.
A {\em discrete} context (or `state') is extracted for each
sample, consisting of a discretised version of some of the most recent
observations preceding it. The set of all relevant contexts are
represented as a discrete context tree. At the bottom level,
a different {\em real-valued} time series model
is associated with each context-state, i.e., with each leaf of the tree.
This defines a very general framework that can be used in conjunction
with any existing model class to build
flexible and {\em interpretable}
mixture models.
Extending the idea of context-tree weighting
leads to 
algorithms that allow
for efficient, \textit{exact} Bayesian inference in this setting. 
The utility of the general framework is illustrated in detail
when autoregressive~(AR) models are used
at the bottom level, resulting in a nonlinear AR mixture model.
The associated 
methods are found to outperform several state-of-the-art
techniques on simulated and real-world~experiments.
\end{abstract}

\section{Introduction}

Modelling and inference of real-valued time series
are critical tasks
with important applications throughout the sciences and engineering. 
A wide range of approaches exist, including classical statistical 
methods~\cite{box,hyndman2008forecasting} 
as well as modern machine learning (ML) techniques, notably  matrix factorisations~\cite{yu2016temporal,faloutsos2018forecasting}, Gaussian processes~\cite{gp,roberts2013gaussian,frigola2015bayesian}, and neural networks~\cite{benidis2020neural,alexandrov2019gluonts,zhang1998forecasting}. Despite their popularity,
there has not been conclusive evidence that in general the latter outperform the former in the time series setting~\cite{makridakis2018statistical,makridakis2018m4,ahmed2010empirical}. Motivated in
part by the two well-known limitations of neural network models (see, e.g.,~\cite{benidis2020neural}), namely, their lack of interpretability 
and their typically very large training data 
requirements, in this work we propose a general class of flexible hierarchical Bayesian models, which are both naturally {\em interpretable} and suitable for applications with limited training data. 
Also, we provide computationally efficient (linear complexity) algorithms for inference and prediction, offering another important practical advantage 
over ML methods~\cite{makridakis2018statistical}.

The first step in the modelling design
is the identification of meaningful discrete states.
Importantly, these are {\em observable} rather than hidden,
and given by the discretised values of some of the most recent samples. 
The second step is the assignment 
of a different time series model to each of these
discrete {\em context-based states}.
In technical terms, we define a hierarchical Bayesian model, 
which at the top level selects the set of relevant states 
(that can be viewed as providing an adaptive partition 
of the state space),
and at the bottom level associates an arbitrary time series 
model to each state. These collections of states
(equivalently, the corresponding state space partitions)
are naturally represented as discrete 
\textit{context-trees}~\cite{ris83a,our}, 
which are shown to admit a natural {\em interpretation} and to enable capturing important aspects of the structure present in the data.
We call the resulting model class, 
the {\em Bayesian Context Trees State Space Model}~(BCT-SSM). 

Although BCT-SSM is referred to as a `model', it is in fact
a general framework for building Bayesian 
mixture models for time series, that can be used in conjunction 
with any existing model class. The resulting model family is rich, 
 and much more general than the class one starts with. 
For example, using any of the standard linear families (like the 
classical AR or ARIMA)
leads to much more general models that
can capture highly nonlinear trends in the data, and 
are easily~interpretable. 

It is demonstrated that
employing this particular type of observable state 
process (as opposed to a conventional hidden state process),
also facilitates very 
effective Bayesian inference.
This is achieved by exploiting the structure of context-tree models
and extending the ideas of context-tree weighting (CTW)~\cite{ctw} and the Bayesian Context Trees~(BCT) framework of~\cite{our}, which were previously 
used only in the restricted setting of \textit{discrete-valued} time series. In this discrete setting, CTW has been used very widely for data compression~\cite{ctw,ctw2}, and BCTs have been used for a range of statistical tasks, including model selection, prediction, entropy estimation, and change-point detection~\cite{ourisit,changepoint,branch_arxiv,lungu2022bayesian,branch_isit,our_entropy}. Context-trees in data compression were also recently studied in~\cite{matsushima2009reducing,nakahara2022probability} and were employed in an optimisation setting in~\cite{ryu2022parameter}.

The resulting tools developed for real-valued data
make the BCT-SSM a powerful Bayesian framework, which
in fact allows for \textit{exact} and computationally very 
efficient Bayesian inference.
In particular, the \textit{evidence}~\cite{mackay} can be computed 
exactly, with all models and parameters integrated out. 
Furthermore, the \textit{a posteriori} most likely~(MAP) partition 
(i.e., the MAP set of discrete states)
can be identified, along with its exact posterior probability.
It is also shown that these algorithms allow 
for efficient sequential updates, which are ideally suited for online 
forecasting, offering an important practical advantage over standard 
ML approaches.   

To illustrate the application of the general framework, 
the case where AR models are used as a building block for the BCT-SSM
is examined in detail.
We refer to the resulting model class as the {\em Bayesian context tree autoregressive} (BCT-AR)~model.
The BCT-AR model
is shown to be a flexible, nonlinear mixture of AR models which is found to outperform several state-of-the-art methods in experiments with 
both simulated and real-world data from standard applications 
of nonlinear time series from economics and~finance, both in terms 
of forecasting accuracy and computational requirements. 

Finally, we note that a number of earlier approaches 
employ discrete patterns in the analysis of~real-valued 
time series~\cite{alvarez2010energy,alvisi2007short,kozat2007universal,  berndt1994using, fu2007stock, hu2014pattern, liu2011novel,  ouyang2010ordinal, hero}. 
These works illustrate the fact that useful and~meaningful information
can indeed be extracted from discrete contexts. However,
in most cases the methods are either application-
or task-specific, and typically resort to {\em ad hoc} considerations
for performing inference. In contrast, in this work,
discrete contexts are used 
in a natural manner by defining
a~hierarchical Bayesian modelling structure upon which principled Bayesian inference is~performed.



\section{The Bayesian Context Trees State Space Model} \label{bct}


\subsection{Discrete contexts} \label{21}


Consider a real-valued time series $(x_1,x_2,\ldots)$.
The first key element of the present development is the use of an 
{\em observable} state 
for each $x_n$, based on discretised versions of some of
the samples $(\ldots,x_{n-2},x_{n-1})$ preceding it. We refer to 
the string consisting of these discretised previous samples as the 
{\em discrete context}; it plays the role of a discrete-valued 
feature vector that can be used to identify useful nonlinear structure 
in the data. These contexts are extracted via 
simple 
quantisers $Q:\mathbb{R}\to A := \{0,1,\dots ,  m-1 \}$, 
of the form,
\begin{equation}
\label{quant}
\hspace*{-0.05 cm } Q(x) \!=\! \left\{
\begin{array}{ll}
\hspace*{-0.05 cm } 0, \ \;\; \;\; \;\;\; x < c_1, \\
\hspace*{-0.05 cm }  i , \ \;\; \;\; \;\;\;\; c_{i} \leq x \leq c _{i+1} ,  \   1\leq i\leq m-2 ,\\
\hspace*{-0.05 cm } m-1, \   x > c_{m-1},
\end{array}
\right.  
\end{equation}
where in this section the thresholds $\{ c_1,\ldots , c_{m-1} \}$ and the resulting quantiser $Q$ are considered fixed.
A systematic way to infer the thresholds from data is described in Section~\ref{hyp}.


This general framework can be used in conjunction with an 
arbitrary way of extracting discrete features, based on
an arbitrary mapping 
to a discrete alphabet, not necessarily of the form in~(\ref{quant}). 
However, the quantisation needs to be meaningful in order to lead to useful 
results. Quantisers as in~(\ref{quant}) offer a generally reasonable choice
although, depending on the application at hand, there are other useful 
approaches, e.g.,~ quantising the percentage differences between 
samples.

\subsection{Context trees}

\begin{wrapfigure}{r}{0.45\linewidth}
\vspace*{-0.35 cm}
  \begin{center}
\vspace*{-0.6 cm}
    \includegraphics[width= 0.8 \linewidth, height= 0.47 \linewidth ]{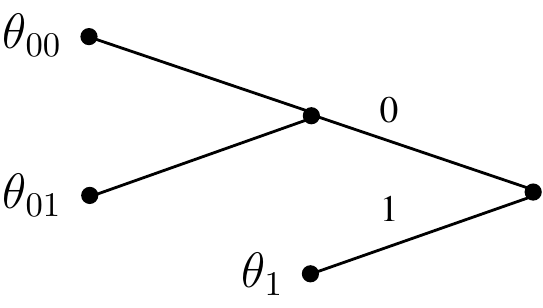}
  \end{center}
\vspace*{-0.4 cm}
\caption{Example of a context tree $T$.}
\label{tree}
\vspace*{-0.4 cm}
\end{wrapfigure}

Given a quantiser $Q$ as in~(\ref{quant}),
a maximum context length \mbox{$D\geq 0$}, 
and a proper $m$-ary context tree $T$,
the context (or `state') of each sample $x_n$ is obtained as follows.
Let \mbox{$t=(Q(x_{n-1}),\ldots,Q(x_{n-D}))$} be the discretised 
string of length~$D$ preceding 
$x_n$; the {\em context} $s$ of $x_n$ is  the unique leaf 
of~$T$ that is a suffix of~$t$. 
For example, for the context tree of Figure~\ref{tree}, if $Q(x_{n-1})=0$ 
and $Q(x_{n-2})=1$ then $s=01$, whereas if $Q(x_{n-1})=Q(x_{n-2})=1$ 
then $s=1$. 

The leaves of the tree define the set of discrete 
states in the hierarchical model.
So, for the example  
BCT-SSM of Figure~\ref{tree}, the set of states 
is $\mathcal {S} = \{ 1, 01, 00\}$.
Equivalently, this process can be viewed as defining
a partition of ${\mathbb R}^2$ into three regions 
indexed by the contexts ${\cal S}$ in~$T.$

To complete the specification of the BCT-SSM, a different time series model~$\mathcal {M}_s$ is associated with each leaf~$s$ of the context tree~$T$, giving a different conditional density for~$x_n$. At time $n$, given the context $s$ determined by the past~$D$ samples $(x_{n-1},\ldots,x_{n-D})$, the distribution of $x_n$ is determined by the model~$\mathcal {M}_s$ assigned to $s$.
Parametric models with parameters $\theta_s$ at each leaf $s$
are considered.
Altogether, the BCT-SSM consists of an
$m$-ary quantiser $Q$, an $m$-ary tree~$T$ that defines the set 
of discrete states, and a collection of parameter vectors~$\theta_s$ for 
the parametric models at its leaves.

Identifying $T$ with the collection of its leaves ${\cal S}$, 
and writing $x_i^j$ for the segment $(x_i,x_{i+i},\ldots,x_j)$, the 
likelihood is,
\begin{align*}\label{lik}
\begin{split}
 p(x|\theta,T) = p(x_1^n | T, \theta, x_{-D+1}^0) = \hspace*{-0.1 cm }\prod _{s \in T}\ \hspace*{-0.25 cm } \prod _ { \hspace*{0.2 cm }i \in B_s} \hspace*{-0.15 cm }p(x_i| T, \theta _s , x_{-D+1}^{i-1} )  , 
\end{split} \vspace*{-0.15 cm}
\end{align*} where  $B_s$ is the set of indices $i\in\{1,2,\ldots ,n\}$ such that the context of $x_i$ is $s$, and \mbox{$\theta=\{\theta_s\;;\;s\in T\}$}. 


\subsection{Bayesian modelling and inference} \label{23}

For the top level we consider collections
of states represented by context trees~$T$ in the class
$\mathcal T (D)$ of all proper $m$-ary trees with 
depth no greater than $D$, where $T$ is {\em proper} if any node in $T$ that is not a leaf has 
exactly $m$~children. 

\smallskip

\textbf{Prior structure.}
For the trees $T\in{\mathcal T}(D)$ 
with maximum depth $D\geq 0$
at the top level of the hierarchical model, we use the Bayesian Context Trees (BCT) prior of~\cite{our},  \vspace*{-0.05 cm}
\begin{equation}\label{prior}
\pi(T)=\pi_D(T;\beta)=\alpha^{|T|-1}\beta^{|T|-L_D(T)} \ , 
\end{equation}
where $\beta\in(0,1)$ is a hyperparameter, $\alpha=(1-\beta)^{1/(m-1)}$,
$|T|$ is the number of leaves of $T$,
and $L_D(T)$ is the number of leaves of
$T$ at depth $D$. 
This prior penalises larger trees
by an exponential amount,
a desirable property that 
discourages overfitting. The default value $\beta= 1-2^{-m+1}$~\cite{our}
is used.
Given a tree model $T \in \mathcal T (D)$, an independent prior 
is placed
on each~$\theta_s$, 
so that $\pi (\theta | T) = \prod _ {s \in T} \pi (\theta_s)$.

\smallskip

Typically, the main obstacle in performing 
Bayesian inference 
with a time series $x$ is the computation of 
the normalising constant (or {\em evidence}) $p(x)$
of the posterior distriburion,
\vspace*{-0.05 cm}
\begin{equation}\label{ctw}
p(x)=\sum_{T \in \mathcal T (D)}\pi(T) \int _ {\theta} p(x|T,\theta) \pi (\theta | T) \  d \theta  . 
\end{equation} 
The power of the proposed Bayesian structure is that, although $ \mathcal T (D)$ is enormously rich, consisting of doubly-exponentially many models in $D$, it is actually possible to
perform \textit{exact} Bayesian inference efficiently.
To that end,
we introduce the Continuous Context Tree Weighting (CCTW) algorithm, and the Continuous Bayesian Context Tree (CBCT) algorithm,
generalising the corresponding algorithms for discrete time series
in~\cite{our}. It is shown that CCTW computes the normalising constant $p(x)$ 
exactly (Theorem~\ref{ctwth}), and CBCT identifies the MAP tree model (Theorem~\ref{bctth}).
The main difference 
from the discrete case 
in~\cite{our}, both in the algorithmic descriptions and in the proofs of the theorems (given in Appendix A), is that a new generalised 
form of \textit{estimated probabilities} is introduced and used
in place of their discrete versions. For a time series $x=x_{-D+1}^n$, these are,
\begin{equation}\label{pes}
P_e(s,x) =  \int \prod _{i \in B_s}  p(x_i|T , \theta _s , x_{-D+1}^{i-1}) \ \pi (\theta _ s) \ d\theta_s . 
\end{equation}
Let $x=x_{-D+1}^n$ be a time series, and let $y_i=Q(x_i)$ denote the corresponding quantised~samples.

\medskip

\textbf{CCTW: Continuous context-tree weighting algorithm}

\vspace*{-0.0 cm}

\begin{enumerate}
\item Build the tree $T_{\text{MAX}}$, whose leaves are all the discrete contexts $y_{i-D} ^ {i-1}, \  i=1,2,\ldots,n$. Compute $P_e(s,x)$ as given in~(\ref{pes}) for each node $s$ of $T_{\text{MAX}}$.

 \item Starting at the leaves and proceeding recursively towards the root compute: \vspace*{-0.05 cm} \[
P_{w,s}\!=\!
\left\{
\begin{array}{ll}
P_e(s,x),  \; \; \; \mbox{if $s$ is a leaf,}\\
\beta P_e(s,x)+(1-\beta)\prod_{j=0}^{m-1} P_{w,sj},  \; \; \; \mbox{o/w,}
\end{array}\!\!
\right.\!\! \vspace*{-0.05 cm}
\]  where $sj$ is the concatenation
of context $s$ and symbol~$j$.
\end{enumerate}

\smallskip

\textbf{CBCT: Continuous Bayesian context tree algorithm}

\vspace*{-0.0 cm}

\begin{enumerate}
\item Build the tree $T_{\text{MAX}}$ and compute $P_e(s,x)$ for each node~$s$ of $T_{\text{MAX}}$, as in CCTW.

\item Starting at the leaves and proceeding recursively towards the root compute: \vspace*{-0.05 cm}  \[
 P_{m,s}\!=\!
\left\{
\begin{array}{ll}
P_e(s,x), \;\;\; \mbox{if $s$ is a leaf at depth $D$,}\\
\beta, \;\;\;   \mbox{if $s$ is a leaf at depth $<D$,}\\
 \max\big\{\beta P_e(s,x),
(1-\beta)\prod_{j=0}^{m-1} \hspace*{-0.07 cm } P_{m,sj}\big\},  \ \mbox{o/w.}
\end{array}
\right. 
\] \item Starting at the root and proceeding recursively with its descendants, for each node: If the maximum is achieved by the first term, prune all its descendants from~$T_{\text{MAX}}$.
\end{enumerate}

\begin{theorem}\label{ctwth}
The weighted probability $P_{w,s}$ at the root is exactly the normalising constant $p(x)$ of~{\em (\ref{ctw})}.
\end{theorem}

\begin{theorem}\label{bctth}
The
tree $T^*_1$ produced by 
the CBCT algorithm is the MAP tree~model,
as long as $\beta\geq 1/2$.
\end{theorem}

\vspace*{-0.1 cm}

Even in cases where the integrals in~(\ref{pes}) are not tractable, 
the fact that they are in the form of standard marginal likelihoods makes it possible to compute them approximately using standard methods, e.g.,~\cite{chib1995marginal,chib2001marginal,wood2011fast}. The above algorithms 
can then be used with these approximations as a way of performing approximate inference for the BCT-SSM. However, this is not investigated further
in this work. Instead, the general principle is illustrated via an 
interesting example where the estimated probabilities can be computed explicitly and the resulting mixture model is a flexible nonlinear model of practical interest. This is described in the next section, where AR models ${\cal M}_s$ are associated to each leaf~$s$. We refer to the resulting model as the Bayesian context tree autoregressive (BCT-AR) model, which is just a particular 
instance of the general BCT-SSM.


\section{The BCT Autoregressive Model}\label{bctar}


Here we consider the BCT-SSM
model class where an AR model of order~$p$
is associated to each leaf $s$ of the tree $T$,
\begin{equation}\label{ar}
x_n = \phi _ {s,1} x_{n-1} + \dots + \phi _ {s,p} x_{n-p} + e_n = {\boldsymbol \phi _ s} ^{\text{T}} \ \mathbf{ \widetilde{ x} } _{n-1} + e_n ,
\end{equation}
where $ e_n \sim \mathcal N (0, \sigma _s ^2) , \ \mathbf{ \widetilde{ x} } _{n-1} = (x_{n-1},\dots, x_{n-p})^{\text{T}} $, and $\boldsymbol \phi _ s = ( \phi _ {s,1}, \dots , \phi _ {s,p} ) ^ {\text{T}}$.


The parameters of the model are the AR coefficients and the noise variance, so that $\theta_s = (\boldsymbol \phi _ s , \sigma _ s ^2)$. An inverse-gamma 
prior is used for the noise variance, and a Gaussian prior is placed on
the AR coefficients, so that the joint prior on the parameters is $\pi(\theta _s) =\pi (\boldsymbol \phi _s | \sigma _s ^2) \pi (\sigma _ s ^2)$, with,
\begin{align*} 
&\pi (\sigma _s ^2 ) = \text {Inv-Gamma} (\tau , \lambda), \quad \quad \pi (\boldsymbol \phi _s | \sigma _s ^2 )  = \mathcal N (\mu _o , \sigma_s ^2 \Sigma _o) , 
\end{align*}
where $(\tau, \lambda,  \mu _o ,  \Sigma _o  )$ are the prior hyperparameters.
This prior specification allows the exact computation of the estimated probabilities of~(\ref{pes}), and gives closed-form posterior distributions
for the AR coefficients and the noise variance. 
These are given in Lemmas~1 and~2,
which are proven in Appendix B. 

\smallskip

\noindent \textbf {Lemma~1.}\; 
\textit{For the AR model, the estimated probabilities $P_e(s,x)$ as 
in~{\em (\ref{pes})} are given by},
\begin{equation}
P_e (s,x) = C_s ^ {-1} \ \frac{ \Gamma \left ( \tau + |B_s| / 2\right )  
\ \lambda ^ \tau } {\Gamma (\tau ) \ 
\left ( \lambda +  D_s / 2  \right ) ^ {\tau + |B_s| / 2 } }  , 
\end{equation}
\vspace*{-0.1cm}
\begin{align*}
\mbox{where}\;
&C_s = \sqrt{ {(2 \pi )^{|B_s|  }}  
	\text {det}( I + \Sigma _o S_3 }),   \\
\mbox{and}\;  
&D_s = s_1 +    \mu _o ^ {\text {T}} \Sigma _ o ^{-1}  \mu _o\\
&\hspace{0.38in}
	  -( \mathbf s_2  +    \Sigma _ o ^{-1}  \mu _o )^ {\text {T}} 
	(S_3 + \Sigma _ o ^{-1} ) ^ {-1}  
	( \mathbf s_2  +    \Sigma _ o ^{-1} \mu _o ),  
\end{align*}
 \textit{with the sums $s_1, \mathbf s_2, S_3 $ defined as:}
\begin{equation*}
s_1 = \sum _ {i \in B_s } x_i ^2, \quad \mathbf s_2 = \sum _ {i \in B_s } x_i \ \mathbf{ \widetilde{ x} } _{i-1}, \quad S_3 = \sum _ {i \in B_s } \mathbf{ \widetilde{ x} } _{i-1}  \mathbf{ \widetilde{ x} } _{i-1} ^{\text{T}}  .
\end{equation*}
\textbf {Lemma~2.}\; \textit{Given a tree model $T$, at each leaf $s$, the posterior distributions of the AR coefficients and the noise variance are,}
\begin{align} \label{10}
&\pi (\sigma _ s ^2 | T,x ) =  \text {Inv-Gamma } \left (\tau + |B_s| / {2} , \lambda + {D_s} / {2} \right )  , \\
&\pi (\boldsymbol \phi _s| T,x ) = t _ \nu (\mathbf m _s, P _s)   ,
\end{align}
\textit{where $t _ \nu$ denotes a multivariate $t$-distribution with $\nu$ degrees of freedom. Here, $ \nu = 2 \tau + |B_s| $, and},
\begin{align*}
\mathbf m_s 
& = 
	(S_3 + \Sigma _o ^ {-1}) ^ {-1} 
	(\mathbf s_2 + \Sigma _o ^ {-1}  \mu _ o)  ,  \\
P_s ^ {-1}
& = 
	\frac {2 \tau +|B_s| } {2 \lambda + D_s} (S_3 + \Sigma _o ^ {-1})  .
\end{align*}
\textbf {Corollary.}\; \textit{ The MAP estimators of $\boldsymbol \phi _s $ and $\sigma _ s ^2$ are given by}, 
\begin{equation*}\label{map}
\widehat {\boldsymbol \phi_s} ^{\text {MAP}} = \mathbf m_s \  , \quad  \widehat {{\sigma _ s ^2 }}^ {\text {MAP}} = ( {2 \lambda + D_s}) / ( {2 \tau + |B_s| + 2} )  .
\end{equation*}

\subsection{Computational complexity and sequential updates} \label{compl}


For each symbol $x_i$ in a time series $x_1^n$, exactly $D+1$ nodes of $T_{\text{MAX}}$ need to be updated, corresponding to its contexts of length $0, 1,\ldots , D$.  For each one of these nodes, only the quantities $\{|B_s|, s_1, \mathbf s_2 , S_3\}$ need to be updated, which can be done efficiently by just adding an extra term to each sum. Using these and Lemma~1, the estimated probabilities $P_e (s,x)$ can be computed for all nodes of $T_{\text{MAX}}$.

\newpage

Hence,
the complexity of both algorithms as a function of $n$ and~$D$ is only $\mathcal{O}(nD)$: {\em linear} in the length 
of the time~series and the maximum depth. Therefore,
the present methods are computationally very efficient 
and scale well with large numbers of observations. 
[Taking into account $m$ and $p$ as well, it is easy to see that
the complexity is $\mathcal{O}\left (nD (p^3+m) \right )$.]

The above discussion also shows that, importantly, all algorithms
can be updated {\em sequentially}.
When observing a new sample $x_{n+1}$, only $D+1$ nodes need to be updated, 
which requires $\mathcal {O}(D)$ operations, i.e., $\mathcal {O}(1)$ 
as a function of~$n$. In particular, this implies
that sequential prediction can be performed very efficiently.
Empirical running times in the forecasting experiments show that the
present methods are much more 
efficient than essentially all the alternatives
examined. In fact, the difference is quite large,
especially when comparing with state-of-the-art ML models that require 
heavy training and do not allow for efficient sequential updates, 
usually giving 
empirical running times that are larger by several orders of 
magnitude; see also~\cite{makridakis2018statistical} for a 
relevant review comparing the computational requirements of ML 
versus statistical techniques.

\vspace*{-0.04 cm}

\subsection{Choosing the quantiser and AR order} \label{hyp}


Finally, a principled Bayesian 
approach is introduced for the selection of 
the quantiser thresholds $\{c_i\}$ of~(\ref{quant}) 
and the AR order~$p$. Viewed
as extra parameters on an additional layer above everything else,
uniform priors are placed on $\{c_i\}$ and $p$, 
and Bayesian model selection~\cite{mackay}
is performed 
to obtain their MAP values:
The resulting posterior distribution $p(\{c_i\},p|x)$ 
is proportional to the \textit{evidence}~$p(x|\{c_i\},p)$, which can 
be computed exactly using the CCTW algorithm (Theorem~\ref{ctwth}). 
So, in order to select appropriate values, suitable ranges of possible 
$\{c_i\}$ and $p$ are specified, and the values with the higher evidence 
are selected.
For the AR order we take $1\leq p \leq p_{\text{max}}$ for an 
appropriate $p_{\text{max}}$ ($p_{\text{max}}=5$ in our experiments), 
and for the $\{c_i\}$ we perform a grid search in a reasonable range 
(e.g., between the 10th and 90th percentiles of the~data).

\section{Experimental results} \label{experiments}

The performance of the BCT-AR model is evaluated on simulated
and real-world data from standard applications of nonlinear time 
series in economics and finance. It is compared with 
the most successful previous approaches for these types of applications, 
considering both classical and modern ML methods. Useful resources 
include the \texttt{R} package \texttt{forecast}~\cite{rfor} and the 
Python 
library \mbox{`GluonTS'}~\cite{alexandrov2019gluonts}, 
containing implementations of state-of-the-art classical and ML methods, 
respectively. We briefly discuss the methods used, and refer to 
the packages' documentation and 
for more details on the methods
themselves and the training procedures carried out.
Among classical statistical approaches, we compare with ARIMA
and Exponential smoothing state space~(ETS) models~\cite{hyndman2008forecasting}, with self-excited threshold autoregressive (SETAR) 
models~\cite{tarinit}, and with mixture autoregressive (MAR) models~\cite{mar}.
Among ML techniques, we compare with 
the Neural Network AR (NNAR) model~\cite{zhang1998forecasting}, and with
the most successful RNN-based approach, `deepAR'~\cite{deepar}.

\subsection{Simulated data} \label{sim}

\vspace{-0.05 cm}

First an experiment on simulated data is performed,
illustrating that the present methods are consistent 
and effective with data generated by BCT-SSM models. The context tree 
used is the tree of Figure~\ref{tree}, the quantiser threshold is $c=0$, and the AR order $p=2$. 
The posterior distribution over trees, $\pi(T|x)$,
is examined. On a time series consisting of only $n=100$ observations, the MAP tree identified by the CBCT algorithm is the empty tree corresponding to a single AR model, with posterior probability~99.9\%. This means that the data do not provide sufficient evidence to support a more complex state space partition. With $n=300$ observations, the MAP tree is now the true underlying model, with posterior probability~57\%. With $n=500$ observations, the posterior of the true~model is~99.9\%. So, the posterior indeed concentrates on the true model, indicating that the BCT-AR inferential framework can be very effective even with limited training data. 

\smallskip


\textbf{Forecasting.} The performance of 
the BCT-AR model is evaluated on out-of-sample 1-step ahead forecasts, 
and compared with state-of-the-art approaches in four simulated and three real datasets. The first simulated dataset (\texttt{sim\_1}) is generated by the BCT-AR model used above, and the second (\texttt{sim\_2}) by a BCT-AR model with a ternary tree of depth 2. The third and fourth ones (\texttt{sim\_3} and \texttt{sim\_4}) are generated from SETAR models of orders $p=1$ and $p=5$,
respectively. In each case, the training set consists of the first~50\% of the observations; also, all models are updated at every time-step in the test set. For BCT-AR, the MAP tree with its MAP parameters is used at every time-step, which can be updated efficiently (Section~\ref{compl}). From Table~\ref{t1}, it is observed that the BCT-AR model outperforms the alternatives, and achieves the lowest mean-squared error (MSE) even 
on the two datasets generated from SETAR models. 
As discussed in Section~\ref{compl}, 
the BCT-AR model  also outperforms the alternatives in terms of empirical 
running times.

\vspace{-0.1 cm}

\begin{table}[!h]
  \centering
  \caption{Mean squared error (MSE) of forecasts}
  \vspace{-0.25 cm}
\label{t1}
  \begin{tabular}{lcccccccccc}
\midrule
\hspace* { -0.2 cm} & \hspace* { -0.2 cm} BCT-AR \hspace* { -0.2 cm} & ARIMA  \hspace* { -0.2 cm} & ETS \hspace* { -0.2 cm}  & NNAR \hspace* { -0.2 cm}  & deepAR \hspace* { -0.2 cm}    & SETAR \hspace* { -0.2 cm}  & MAR \hspace* { -0.2 cm}  \\
 \midrule
 \hspace* { -0.2 cm} \texttt {sim\_1} \hspace* { -0.2 cm} & \hspace* { -0.2 cm} \bf 0.131 & 0.150 & 0.178 & 0.143 & 0.148  &  0.141 & 0.151 \hspace* { -0.2 cm} \\
 \hspace* { -0.2 cm} \texttt {sim\_2} \hspace* { -0.2 cm} & \hspace* { -0.2 cm} \bf  0.035 & 0.050 & 0.054 & 0.048 & 0.061 &   0.050 & 0.064  \hspace* { -0.2 cm} \\
\hspace* { -0.2 cm}  \texttt {sim\_3} \hspace* { -0.2 cm} & \hspace* { -0.2 cm} \bf 0.216 & 0.267 & 0.293 & 0.252 & 0.273  &  0.243 & 0.283   \hspace* { -0.2 cm} \\
\hspace* { -0.2 cm}  \texttt {sim\_4} \hspace* { -0.2 cm} & \hspace* { -0.2 cm} \bf 0.891  & 1.556 & 1.614 & 1.287  & 1.573 &   0.951 & 1.543 \hspace* { -0.2 cm} \\
  \midrule
\hspace* { -0.2 cm}  \texttt {unemp} \hspace* { -0.2 cm} & \hspace* { -0.2 cm} \bf 0.034 & 0.040 & 0.042 & 0.036 & 0.036  & 0.038 & 0.037 \hspace* { -0.2 cm} \\
\hspace* { -0.2 cm}  \texttt {gnp} \hspace* { -0.2 cm} & \hspace* { -0.2 cm} \bf 0.324 & 0.364 & 0.378 & 0.393 & 0.473   & 0.394 & 0.384  \hspace* { -0.2 cm} \\
\hspace* { -0.2 cm}  \texttt{ibm} \hspace* { -0.2 cm} & \hspace* { -0.2 cm} 78.02 & 82.90 & 77.52 & 78.90 & \bf 75.71   & 81.07 & 77.02 \hspace* { -0.2 cm} \\
    \midrule
  \end{tabular}
  \vspace*{-0.4 cm}
\end{table}

\subsection{US unemployment rate}

\vspace*{-0.05 cm}

An important application of SETAR models is in modelling the US unemployment rate~\cite{tarec,montg,tsay2005analysis}. As described in~\cite{tsay2005analysis}, the unemployment rate moves countercyclically with business cycles, and rises quickly but decays slowly, indicating nonlinear behaviour. Here, the quarterly US unemployment rate in the time period from 1948 to 2019 is examined (dataset \texttt{unemp}, 288 observations). Following~\cite{montg}, the difference series 
\mbox{${\Delta x}_n = x_n-x_{n-1}$} is considered, and a constant term 
is included in the AR model. 
For the quantiser alphabet size, $m=2$ is a natural choice, as will become apparent below. The threshold selected using the procedure of Section~\ref{hyp} is $c=0.15$,
and the MAP tree is the tree of Figure~\ref{tree}, with depth $d=2$, leaves~$\{1, 01, 00 \}$, and posterior
probability 91.5\%. The fitted BCT-AR model with its MAP parameters is,
\[ 
{\Delta x}_n \!=\! \left\{
\begin{array}{ll}
 0.09 +  0.72   {\Delta x}_{n-1} - 0.30   {\Delta x}_{n-2} + 0.42 \ e_n,   \\
0.04 + 0.29  \ {\Delta x}_{n-1} - 0.32  \ {\Delta x}_{n-2} + 0.32 \ e_n, \\
\hspace*{-0.25 cm} -0.02 + 0.34 \ {\Delta x}_{n-1} + 0.19  \ {\Delta x}_{n-2}+ 0.20 \ e_n, 
\end{array}
\right.  \vspace*{-0.15 cm}
\] 
with $e_n \sim \mathcal N (0, 1)$, and corresponding regions $s = 1$ if $ {\Delta x}_{n-1}>0.15$, $s = 01$ if ${\Delta x}_{n-1}\leq 0.15, {\Delta x}_{n-2}>0.15$, and $s = 00$ if ${\Delta x}_{n-1}\leq 0.15, \  {\Delta x}_{n-2}\leq 0.15 $.

\smallskip

\textbf {Interpretation}. The MAP BCT-AR model finds significant structure in the data, providing a very natural interpretation. It identifies 3 meaningful states: First, jumps in the unemployment rate higher than 0.15 signify economic contractions (context 1). If there is no jump at the most recent time-point, the model looks further back to determine the state. Context 00 signifies a stable economy, as there are no jumps in the unemployment rate for two consecutive quarters. Finally, context~01 identifies an intermediate state: ``stabilising just after a contraction''. An important feature identified by the BCT-AR model is that the volatility is different in each case: Higher in contractions ($\sigma = 0.42$), smaller in stable economy regions ($\sigma = 0.20$), and in-between for context~01 ($\sigma = 0.32$). 

\smallskip

\textbf{Forecasting.} In addition to its appealing interpretation, the BCT-AR model outperforms all benchmarks in forecasting, giving a 6\% lower MSE than the second-best~method~(Table~\ref{t1}).

\subsection{US Gross National Product}

\begin{wrapfigure}{r}{0.42\linewidth}
  \begin{center}
  \vspace*{-1.0 cm}
    \includegraphics[width= 0.85 \linewidth, height= 0.45 \linewidth ]{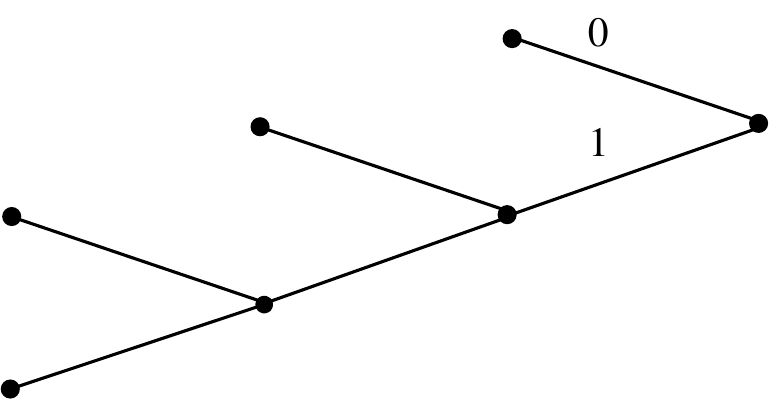}
  \end{center}
\vspace*{-0.3 cm}
\caption{MAP tree}
\label{tree_gnp}
\vspace*{-0.3 cm}
\end{wrapfigure}

Another important example of a nonlinear time series is the US Gross National Product~(GNP)~\cite{potter1995nonlinear,tarec}. 
The quarterly US GNP in the time period from 1947 to 2019 is examined
(291 observations, dataset \texttt{gnp}). Following~\cite{potter1995nonlinear}, the difference in the logarithm of the series, 
$y_n = \log x_n - \log x_{n-1}$, is considered.
As above, $m=2$ is a natural choice for the quantiser size, helping to differentiate economic expansions from contractions -- which govern the underlying dynamics. The MAP BCT-AR tree is shown in Figure~\ref{tree_gnp}: It has depth $d=3$, its leaves are $\{0, 10, 110, 111 \}$ and its posterior 
probability is~42.6\%.

\smallskip

\textbf{Interpretation.} Compared with the previous example, here the 
MAP BCT-AR model finds even richer structure in the data and 
identifies four meaningful states. First, as before, there is a single 
state corresponding to an economic contraction (which now corresponds 
to $s=0$ instead of $s=1$, as the GNP obviously increases in expansions 
and decreases in contractions). And again, the model does not look further back whenever a contraction is detected. Here, the model also shows that the effect of a contraction is still present even after {\em three} quarters ($s=110$),
and that the exact `distance' from a contraction is also important, with
the dynamics changing depending on how much time has elapsed.
Finally, the state $s=111$ corresponds to a flourishing, expanding economy, 
without a contraction in the recent past. An important feature 
captured by the BCT-AR model is again that the volatility is different in each 
case. More specifically, it is found that the volatility strictly decreases with the 
distance from the last contraction, starting with the maximum 
$\sigma =1.23$ for $s=0$ and decreasing to $\sigma = 0.75$ 
for $s=111$.

\smallskip

\textbf{Forecasting.} The BCT-AR model outperforms all benchmarks in forecasting, giving a 12\% lower MSE than the second-best method, as presented in Table~\ref{t1}.

\vspace*{-0.05 cm}

\subsection{The stock price of IBM}

\begin{wrapfigure}{r}{0.4\linewidth}
  \begin{center}
  \vspace*{-0.7 cm}
    \includegraphics[width= 0.78 \linewidth, height= 0.4 \linewidth ]{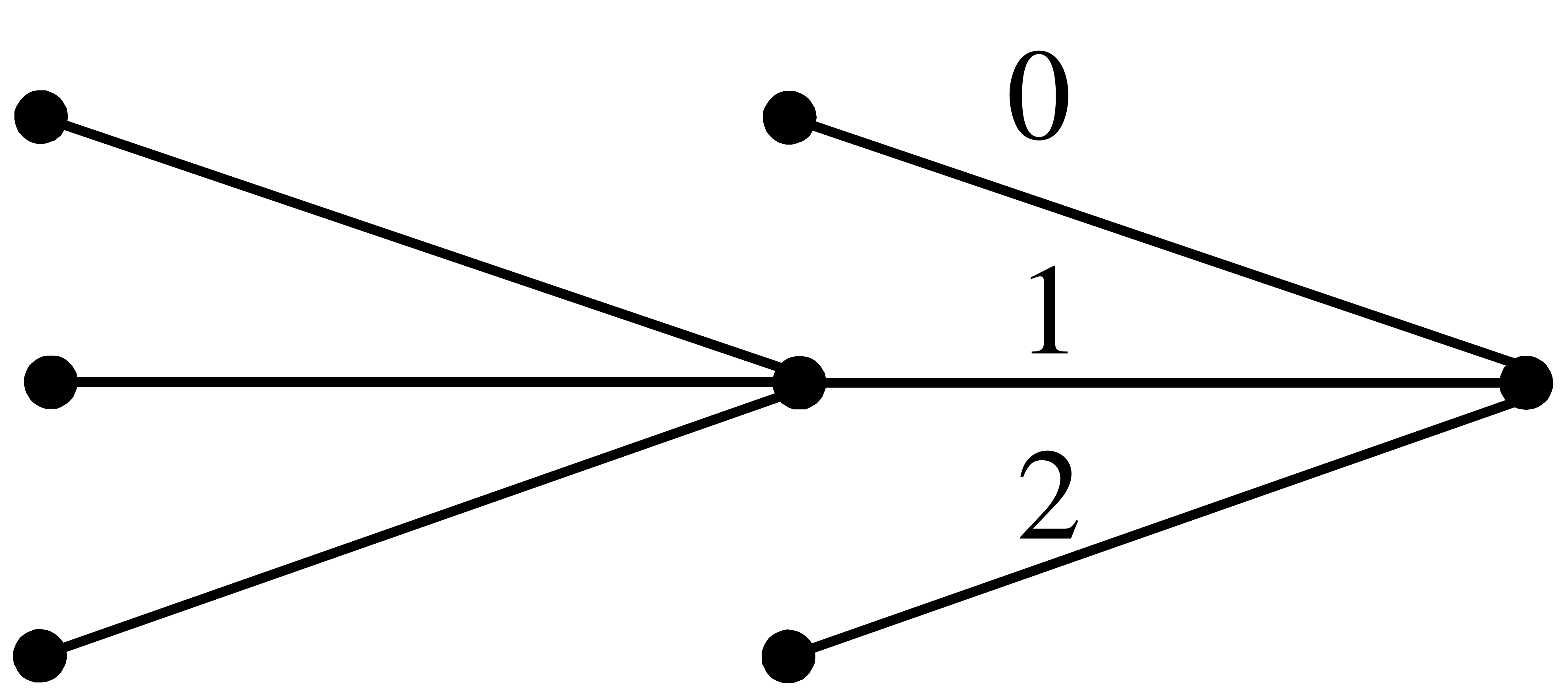}
  \end{center}
\vspace*{-0.25 cm}
\caption{MAP tree}
\label{tree_ibm}
\vspace*{-0.5 cm}
\end{wrapfigure}

Finally, the daily IBM common stock closing price from  May 17, 1961 to  November~2, 1962 is examined (dataset \texttt{ibm}, 369 observations), taken from~\cite{box}. This is a well-studied dataset, with~\cite{box} fitting an ARIMA model,~\cite{tong1990non} fitting a SETAR model, and~\cite{mar} fitting a MAR model to the data. Following previous approaches, the first-difference series, \mbox{${\Delta x}_n = x_n-x_{n-1}$}, is considered. The value $m=3$ is chosen
for  the alphabet size of the quantiser,
with contexts~$\{ 0,1,2\}$ naturally corresponding
to the states \{down, steady, up\} for the stock price.
Using the procedure of Section~\ref{hyp} to select the thresholds, the resulting quantiser regions are: $s=0$ if ${\Delta x}_{n-1}<-7 $, $s=2$ if ${\Delta x}_{n-1}>7$, and $s=1$~otherwise. The MAP tree is shown in Figure~\ref{tree_ibm}: It has depth $d=2$, and its leaves are~$\{0,2,10,11,12 \}$, hence identifying five states. Its posterior probability is 99.3\%, suggesting that there is very strong evidence in the data supporting this exact structure, even with only 369 observations. 

\medskip




\textbf{Interpretation.} The BCT-AR model reveals important information about apparent structure in the data. Firstly, it admits a very simple and natural interpretation: In order to determine the AR model generating the next value, 
one needs to look back until there is a significant enough price change (corresponding to contexts 0, 2, 10, 12), or until the maximum depth of 2 (context~11)
is reached. Another important feature captured by this model is the commonly observed asymmetric response in volatility due to positive and negative shocks, sometimes called the {\em leverage effect}~\cite{tsay2005analysis, box}. Even though there is no suggestion of that in the prior, the MAP parameter
estimates show that negative shocks increase the volatility much more: Context 0 has the highest volatility ($\sigma  =12.3$), with 10 being a close second ($\sigma = 10.8$), showing that the effect of a past shock is still present. In all other cases the volatility is much smaller 
($5.17\leq \sigma\leq 6.86$). 

\smallskip

\textbf{Forecasting.} As shown Table~\ref{t1}, in  this example deepAR performs~marginally better in \mbox{1-step} ahead forecasts, with BCT-AR, ETS and MAR also having comparable performance.

\newpage

\newpage

\onecolumn

\twocolumn

{  
\bibliographystyle{plain}

}

\onecolumn

\newpage 

\appendix





\setcounter{section}{1}

\section*{A. Proofs of Theorems~1 and~2} \label{appA}

The important observation here is that using the new, different form of the 
estimated probabilities $P_e(s,x)$ in~(\ref{pes}), it is still
possible to factorise the marginal likelihoods $p(x|T)$ for the general BCT-SSM as,
\begin{equation*}
p(x|T)
=\int p(x|\theta,T)\pi(\theta|T)  d\theta =\int  \prod _{s \in T} \bigg (  \prod _ {i \in B_s} p(x_i| T, \theta _s , x_{-D+1}^{i-1} )  \ \pi (\theta_s) \  d\theta_s \bigg )  
=\prod_{s\in T} P_e(s,x),
\end{equation*}
where the second equality follows from the likelihood of the general BCT-SSM, and the fact that we use independent priors on the parameters at the leaves, so that $\pi(\theta | T) = \prod _ {s  \in T }\pi (\theta _ s)$. 
Then, the proofs of Theorems~1 and~2 follow along the same lines as the proofs of the corresponding results for discrete time series in~\cite{our}, with the main difference being that the new general version of the estimated probabilities $P_e(s,x)$ of~(\ref{pes}) need to be used in place of their discrete versions.

The $k$-BCT algorithm of~\cite{our} can be generalised in a similar manner to the way the CTW and BCT algorithms were generalised.
The resulting algorithm identifies the top-$k$ {\em a posteriori} most likely context trees. The proof of the theorem claiming this is similar to the proof of Theorem~3.3 of~\cite{our}. Again, the important difference, both in the algorithm description and in the proof, is that the estimated probabilities $P_e(s,x)$ are used in place of their discrete counterparts $P_e(a_s)$.

\setcounter{section}{2}

\section*{B. Proofs of Lemmas~1 and~2} \label{appB}

The proofs of these lemmas are mostly based on explicit computations. 
Recall that, for each context~$s$, the set~$B_s$ consists of those indices $i\in\{1,2,\ldots ,n\}$ such that the context of $x_i$ is $s$. The important step in the following two proofs is the factorisation of the likelihood using the sets $B_s$.
In order to prove the lemmas for the AR model with parameters $\theta_s = (\boldsymbol \phi _ s , \sigma _ s ^2)$, we first consider an intermediate step 
in which the noise variance is assumed to be known and equal to~$\sigma ^2$.

\subsection*{B.1. Known noise variance}

Here, 
an AR model with known variance $\sigma ^2$ is associated with
every leaf $s$ of the context tree $T$, 
so that,
\begin{equation}\label{ar_2}
x_n = \phi _ {s,1} x_{n-1} + \dots + \phi _ {s,p} x_{n-p} + e_n = {\boldsymbol \phi _ s} ^{\text{T}} \ \mathbf{ \widetilde{ x} } _{n-1} + e_n , \quad  e_n \sim \mathcal N (0, \sigma ^2 ).
\end{equation}
In this setting, the parameters of the model are only the AR coefficients: $\theta_s \hspace*{-0.08 cm}= \hspace*{-0.03 cm} \boldsymbol \phi _ s$.\hspace*{-0.01 cm} For these, a Gaussian~prior is used, 
\begin{equation}
\theta _s\sim \mathcal N ( \mu _o , \Sigma _ o) \ ,
\end{equation}
where $\mu _o , \Sigma _ o$ are  hyperparameters. 

\smallskip

{Lemma~B.} \textit{The estimated probabilities  $P_e(s,x)$ for the known-variance case are given by,}
\begin{equation} \label{ar_pes_known}
P_e(s,x) = \frac{1}{(2 \pi \sigma ^2)^{|B_s| /2 }} \ \frac{1}{\sqrt{\text {det}( I + \Sigma _o S_3 / \sigma ^2})}  \  \exp{\bigg \{ - \frac {E_s} {2 \sigma ^2}\bigg \} },
\end{equation}
\textit{where $I$ is the identity matrix and $E_s$ is given by: }
\begin{equation}\label{ar_pes_known2}
E_s = s_1 + \sigma ^2   \mu _o ^ {\text {T}} \Sigma _ o ^{-1} \mu _o  - ( \mathbf s_2  +  \sigma ^2  \Sigma _ o ^{-1} \mu _o )^ {\text {T}} (S_3 + \sigma ^ 2\Sigma _ o ^{-1} ) ^ {-1}  ( \mathbf s_2  +  \sigma ^2  \Sigma _ o ^{-1} \mu _o )  \ .
\end{equation}

\begin{proof}
For the AR model of~(\ref{ar_2}),
$$ p(x_i|T , \theta _s , x_{-D+1}^{i-1}) = \frac{1}{\sqrt{2 \pi \sigma ^2}} \  \exp \bigg \{ -\frac {1}{2 \sigma ^2} (x_i - {\theta _ s} ^{\text{T}} \mathbf{ \widetilde{ x} } _{i-1} ) ^2 \bigg \},$$
so that,
$$\prod _{i \in B_s}  p(x_i|T , \theta _s , x_{-D+1}^{i-1}) = \frac{1}{(\sqrt{2 \pi \sigma ^2})^{|B_s|}} \  \exp \bigg \{ -\frac {1}{2 \sigma ^2} \sum _{i \in B_s}(x_i -  {\theta _ s} ^{\text{T}} \mathbf{ \widetilde{ x} } _{i-1} ) ^2 \bigg \}.
$$
Expanding the sum in the exponent gives,
\begin{eqnarray*}
\sum _{i \in B_s}(x_i - {\theta _ s} ^{\text{T}} \mathbf{ \widetilde{ x} } _{i-1} ) ^2 
=
    \sum _{i \in B_s} x_i ^2 - 2  \theta_s ^{\text{T}} \sum _{i \in B_s} x_i \mathbf{ \widetilde{ x} } _{i-1} +  \theta_s ^{\text{T}} \sum _ {i \in B_s } \mathbf{ \widetilde{ x} } _{i-1}  \mathbf{ \widetilde{ x} } _{i-1} ^{\text{T}}  \theta_s
=
    s_1 - 2  \theta_s ^{\text{T}} \mathbf s_2 +  \theta_s ^{\text{T}} S_3  \theta_s, 
\end{eqnarray*}
from which we obtain that,
\begin{align*}
\prod _{i \in B_s}  p(x_i|T , \theta _s , x_{-D+1}^{i-1}) = \frac{1}{(\sqrt{2 \pi \sigma ^2})^{|B_s|}} \  \exp \bigg \{ -\frac {1}{2 \sigma ^2} (s_1 - 2  \theta_s ^{\text{T}} \mathbf s_2 +  \theta_s ^{\text{T}} S_3  \theta_s  )  \bigg \}  
 =  (\sqrt {2 \pi }) ^ p  \rho_s \ \mathcal N ( \theta_s ;\boldsymbol \mu , S )  ,
\end{align*}
by completing the square, where  $\boldsymbol \mu = S_3^{-1} \mathbf s_2$,  $ S = \sigma ^2 S_3 ^{-1}$, and,
\begin{equation}
\rho _s =\sqrt{ \frac{\text{det}(\sigma^2  S_3 ^{-1})}{(2 \pi \sigma ^2)^{|B_s|}}} \  \exp \bigg \{  -\frac {1}{2 \sigma ^2} (s_1 - \mathbf s_2 ^{\text{T}} S_3 ^{-1} \mathbf s_2) \bigg \}  .
\end{equation}

\newpage

\noindent So, multiplying with the prior,
\begin{equation*}
\prod _{i \in B_s}  p(x_i|T , \theta _s , x_{-D+1}^{i-1}) \pi ( \theta _ s)= (\sqrt {2 \pi }) ^ p  \rho_s \ \mathcal N ( \theta_s ;\boldsymbol \mu , S ) \ \mathcal N ( \theta _ s;  \mu _o , \Sigma _ o) =  \rho_s Z_s \ \mathcal N ( \theta_s ;\mathbf m , \Sigma ) ,
\end{equation*}
where $ \Sigma^{-1} = \Sigma _ o ^{-1} +  S ^{-1}, \ m = \Sigma \ (\Sigma _ o ^{-1} \mu_o + S^{-1} \boldsymbol \mu)$, and, 
\begin{equation}
Z _s =   \frac{1}{\sqrt{{\text {det} (\Sigma_o} +\sigma^2  S_3 ^{-1}) }} \  \exp{\bigg \{ -   \frac {1}{2} (\mu _o - S_3 ^{-1} \mathbf s_2)^{\text{T}} (\Sigma_o +\sigma^2  S_3 ^{-1}) ^{-1} ( \mu _o - S_3 ^{-1} \mathbf s_2)\bigg \} } \ .
\end{equation}
Therefore,
\begin{equation}
\prod _{i \in B_s}  p(x_i|T , \theta _s , x_{-D+1}^{i-1}) \pi ( \theta _ s)= \rho_s Z_s \ \mathcal N ( \theta_s ;\mathbf m , \Sigma ), \label{13} 
\end{equation}
and hence,
$$
P_e(s,x) =  \int \prod _{i \in B_s}  p(x_i|T , \theta _s , x_{-D+1}^{i-1}) \ \pi (\theta _ s) \ d\theta_s \ = \rho_s Z_s .$$
Using standard matrix inversion properties, after some algebra the product $\rho_s Z_s $ can be rearranged to give exactly the required expression 
in~(\ref{ar_pes_known}).
\end{proof}

\subsection*{B.2. Proof of Lemma~1}

Getting back to the original case as described in the main text, 
the noise variance is considered to be a parameter of the AR model,
so that,
$\theta_s = (\boldsymbol \phi _ s , \sigma _ s ^2)$. Here, the joint prior on the parameters is \mbox{$\pi(\theta _s) =\pi (\boldsymbol \phi _s | \sigma _s ^2) \pi (\sigma _ s ^2)$}, where, \vspace*{-0.1 cm}
\begin{align} \label{ar_pr_ap}
\sigma _s ^2\sim \mbox{Inv-Gamma}(\tau , \lambda)  , \; \; 
\boldsymbol \phi _s | \sigma _s ^2 \sim \mathcal N (\mu _o , \sigma_s ^2 \Sigma _o)  ,
\end{align}
and where $(\tau, \lambda,  \mu _o ,  \Sigma _o  )$ are hyperparameters.
For $P_e(s,x)$, we just need to compute the integral:
\begin{align}
P_e(s,x) =  \int \prod _{i \in B_s}  p(x_i|T , \theta _s , x_{-D+1}^{i-1}) \ \pi (\theta _ s ) \ d\theta_s  
= \int \pi (\sigma _ s ^2) \left ( \int  \prod _{i \in B_s}  p(x_i|T , \boldsymbol \phi _ s , \sigma _ s ^2 , x_{-D+1}^{i-1}) \ \pi (\boldsymbol \phi _ s| \sigma _ s ^2) \ d\boldsymbol \phi _ s\right ) d\sigma_s ^2 .
\end{align}
The inner integral has exactly the form of the estimated probabilities $P_e(s,x)$ from the previous section, where the noise variance was fixed. The only difference is that the prior $\pi (\boldsymbol \phi _ s | \sigma _ s ^2)$ of~(\ref{ar_pr_ap}) now has covariance matrix~$\sigma _ s ^2 \Sigma _o $ instead of $\Sigma _o$. So, using~(\ref{ar_pes_known})-(\ref{ar_pes_known2}), with $\Sigma _o$ replaced by $\sigma _ s ^2 \Sigma _o $, we~get, 
\begin{equation*}
P_e(s,x) =\int \pi (\sigma _ s ^2) \bigg \{ C_s ^ {-1}\bigg (\frac{1}{\sigma _ s ^2}  \bigg ) ^ {{|B_s|}/{2}} \exp \bigg ( - \frac{D_s}{2 \sigma _ s ^2} \bigg )  \bigg \} d\sigma_s ^2, 
\end{equation*}
with $C_s$ and $D_s$ as in Lemma~1. And using the inverse-gamma prior $\pi (\sigma _ s ^2)$ of~(\ref{ar_pr_ap}), 
\begin{equation}\label{19}
P_e(s,x)= \ C_s ^ {-1} \  \frac {\lambda ^ {\tau}}{\Gamma (\tau)} \  \int  \bigg (\frac{1}{\sigma _ s ^2}  \bigg ) ^ {\tau ' +1 }  \exp \bigg ( - \frac{\lambda '}{ \sigma _ s ^2} \bigg )  d\sigma_s ^2 , 
\end{equation}
with $\tau ' = \tau + \frac{|B_s|}{2} $ and $\lambda ' = \lambda + \frac{D_s}{2}$.
The integral in~(\ref{19}) has the form of an inverse-gamma density with parameters $\tau ' $ and~$\lambda '$. Its closed-form solution, as required, completes the proof of the lemma:
\[
P_e(s,x) =  C_s ^ {-1} \  \frac {\lambda ^ {\tau}}{\Gamma (\tau)} \ \frac {\Gamma (\tau ' )} {\left ( \lambda ' \right )^ {\tau '}}   . 
\vspace*{-0.4 cm} \]  \qed

\subsection*{B.3. Proof of Lemma~2}

In order to derive the required expressions for the posterior distributions of $\boldsymbol \phi _ s $ and $\sigma _ s ^2$ , for a leaf $s$ of model $T$, first consider the joint posterior $\pi (\theta _ s | T, x) = \pi (\boldsymbol \phi _ s , \sigma _ s ^2 | T, x)$, given by,
\begin{align*}
\pi (\theta _ s | T, x) \propto  p(x| T, \theta _ s )   \pi (\theta_ s )   =    \prod _{i=1} ^ n  p(x_i|T , \theta _s ,  x_{-D+1}^{i-1})       \pi (\theta_ s ) 
 \propto \prod _{i \in B_s}   p(x_i|T , \theta _s ,  x_{-D+1}^{i-1})       \pi (\theta_ s ) , 
\end{align*} 
where we used the fact that, in the product, only the terms involving indices $i\in B_s$ are functions of~$\theta _s$. So,
\begin{align*}
\pi (\boldsymbol \phi _ s , \sigma _ s ^2 | T, x)  &\propto \left ( \  \prod _{i \in B_s}   p(x_i|T ,\boldsymbol \phi _ s , \sigma _ s ^2 , x_{-D+1}^{i-1})      \  \pi (\boldsymbol \phi _ s | \sigma _ s ^2)  \right ) \pi (\sigma _ s ^2 )  .
\end{align*}
Here, the first two terms can be computed from~(\ref{13}) of the previous section, where the noise variance was known. Again, the only difference is that we have to replace $\Sigma _o $ with $\sigma _ s ^2 \Sigma _o $ because of the prior $ \pi (\boldsymbol \phi _ s | \sigma _ s ^2)$ defined in~(\ref{ar_pr_ap}). After some algebra,  
\begin{align*}
\pi (\boldsymbol \phi _ s , \sigma _ s ^2 | T, x)  &\propto  \bigg (\frac{1}{\sigma _ s ^2}  \bigg ) ^ {{|B_s|}/{2}} \exp \bigg ( - \frac{D_s}{2 \sigma _ s ^2} \bigg ) \  \mathcal N ( \boldsymbol \phi _ s ;\mathbf m_s , \Sigma _s )   \ \pi (\sigma _ s ^2 ) \ ,
\end{align*}
with $\mathbf m _s $  defined as in Lemma~2, and $\Sigma _ s = \sigma _ s  ^2 (S_3 + \Sigma _ o ^{-1}) ^{-1} $. 
 Substituting the prior $\pi (\sigma _ s ^2)$ in the last expression,
\begin {equation}\label{jointpost}
\pi (\boldsymbol \phi _ s , \sigma _ s ^2 | T, x)  \propto  \bigg (\frac{1}{\sigma _ s ^2}  \bigg ) ^ {\tau + 1 + {|B_s|}/{2}} \exp \bigg ( - \frac{\lambda + D_s/2}{\sigma _ s ^2} \bigg ) \  \mathcal N ( \boldsymbol \phi _ s ;\mathbf m_s , \Sigma _s )   . 
\end{equation}
From~(\ref{jointpost}), it is easy to integrate out $\boldsymbol \phi _ s $ and get the posterior of $\sigma _s ^2$,
\[
\pi (\sigma _ s ^2 | T , x) = \int \pi (\boldsymbol \phi _ s , \sigma _ s ^2 | T, x) \ d \boldsymbol \phi _ s \propto  \bigg (\frac{1}{\sigma _ s ^2}  \bigg ) ^ {\tau + 1 + {|B_s|}/{2}} \exp \bigg ( - \frac{\lambda + D_s/2}{\sigma _ s ^2} \bigg ),
\]
which is of the form of an inverse-gamma distribution with parameters  $\tau ' = \tau + \frac{|B_s|}{2} $ and $\lambda ' = \lambda + \frac{D_s}{2}$, proving the first part of the lemma.

\smallskip

\noindent However, as $\Sigma _ s$ is a function of $\sigma _ s ^ 2$, integrating out $\sigma_s^2$ requires more algebra.
We have,
\begin{align*}
\mathcal N ( \boldsymbol \phi _ s ;\mathbf m_s , \Sigma _s ) &\propto \frac {1}{\sqrt {\text {det} (\Sigma _s)}} \ \exp \bigg \{  -\frac {1}{2} (\boldsymbol \phi _ s -\mathbf m _ s) ^ {\text {T}} \Sigma _ s^ {-1} (\boldsymbol \phi _ s-\mathbf m _ s)  \bigg \}  \\
& \propto \bigg ( \frac {1} {\sigma _ s ^2} \bigg ) ^ {p/2}  \exp \bigg \{  -\frac {1}{2\sigma _ s ^2} (\boldsymbol \phi _ s-\mathbf m _ s) ^ {\text {T}} (S_3 + \Sigma _o ^{-1})  (\boldsymbol \phi _ s -\mathbf m _ s)  \bigg \} \ ,
 \end{align*}
and substituting this in~(\ref{jointpost}) gives
that $\pi (\boldsymbol \phi _ s  , \sigma _ s ^2 | T, x)$ is proportional to,
\begin{align*}
\bigg (  \frac {1} {\sigma _ s ^2}\bigg ) ^ {\tau +1 + \frac {|B_s|+ p} {2}} \hspace{-0.1 cm} \exp  \bigg \{  -\frac {1}{2\sigma _ s ^2}\bigg ( 2 \lambda + D_s + (\boldsymbol \phi _ s  -\mathbf m _ s) ^ {\text {T}} (S_3 + \Sigma _o ^{-1})  (\boldsymbol \phi _ s  -\mathbf m _ s) \bigg ) \bigg \} ,
 \end{align*}
which, as a function of $\sigma_s ^2$, has the form 
of an inverse-gamma density, allowing $\sigma^2$ to be integrated
out. Denoting \mbox{$L =   2 \lambda + D_s + (\boldsymbol \phi _ s-\mathbf m _ s) ^ {\text {T}} (S_3 + \Sigma _o ^{-1})  (\boldsymbol \phi _ s -\mathbf m _ s)$}, and $ \widetilde \tau = \tau +\frac {|B_s|+ p} {2}$, 
\begin{align*}
&\pi (\boldsymbol \phi _ s | T , x) = \int \pi (\boldsymbol \phi _ s  , \sigma _ s ^2 | T, x) \ d  \sigma _ s ^2 \propto \int  \bigg (  \frac {1} {\sigma _ s ^2}\bigg ) ^ {\widetilde \tau +1 }  \exp \bigg ( - \frac{L } {2 \sigma _ s ^2}\bigg ) \ d \sigma _ s ^2 = \frac {\Gamma (\widetilde \tau)} {(L/2 ) ^ {\widetilde \tau}} \ .
 \end{align*}
So, as a function of $\boldsymbol \phi _ s $, the posterior $\pi(\boldsymbol \phi _ s |T,x)$ is,
\begin{align*}
\pi (\boldsymbol \phi _ s | T , x) \propto L^ {- \widetilde \tau} &= \bigg (  2 \lambda + D_s + (\boldsymbol \phi _ s -\mathbf m _ s) ^ {\text {T}} (S_3 + \Sigma _o ^{-1})  (\boldsymbol \phi _ s-\mathbf m _ s) \bigg ) ^ {- \frac {2 \tau + |B_s| + p}{2}} \\
& \propto \bigg ( 1 + \frac {1}{2 \tau + |B_s|} \  (\boldsymbol \phi _ s-\mathbf m _ s) ^ {\text {T}}\frac {(S_3 + \Sigma _o ^{-1})(2 \tau + |B_s|)}{(2 \lambda + D_s) }  (\boldsymbol \phi _ s -\mathbf m _ s)  \bigg ) ^ {- \frac {2 \tau + |B_s| + p}{2}} \\
& \propto \bigg ( 1 + \frac {1}{\nu} \  (\boldsymbol \phi _ s -\mathbf m _ s) ^ {\text {T}} P_s ^{-1}  (\boldsymbol \phi _ s -\mathbf m _ s)  \bigg ) ^ {- \frac {\nu+ p}{2}}  ,
\end{align*}
which is exactly in the form of a multivariate $t$-distribution, with $p$ being the dimension of $\boldsymbol \phi _ s$, and with $\nu, \mathbf m _ s$ and  $P _s $ exactly as given in Lemma~2, completing the proof. \qed

\newpage

\end{document}